\input{psfig}
\documentstyle [12pt,twoside,epsf]{article}

\def\unlock{\catcode`@=11} 
\def\lock{\catcode`@=12} 
\unlock
  \def\leftrightarrowfill{$\m@th \mathord\leftarrow \mkern-6mu
     \cleaders\hbox{$mkern-2mu \mathord- \mkern-2mu$}\hfill
     \mkern-6mu \mathord\rightarrow$}
  \def\overleftrightarrow#1{\vbox{\ialign{##\crcr
     \leftrightarrowfill\crcr\noalign{\kern-1pt\nointerlineskip}
     $\hfil\displaystyle{#1}\hfil$\crcr}}}
\lock

\newcommand{\bq}{\begin{equation}}
\newcommand{\ba}{\begin{eqnarray}}
\newcommand{\eq}{\end{equation}}
\newcommand{\ea}{\end{eqnarray}}
\newcommand{\half}{{1\over 2}}

\def\b{\beta}

\def\d{\delta}
\def\e{\epsilon}
\def\f{\phi}
\def\vf{\varphi}

\def\j{\psi}

\def\q{\theta}
\def\r{\rho}

\def\F{\Phi}

\def\J{\Psi}

\def\S{\Sigma}

\def\cc{{\cal C}}
\def\cd{{\cal D}}

\def\ct{{\cal T}}

\def\sq{{\lower.2ex\hbox{\large$\Box$}}}
\def\pa{\partial}

\def\TH{{\raise.2ex\hbox{$\displaystyle \bigodot$}\mskip-4.7mu \llap H \;}}
\def\face{{\raise.2ex\hbox{$\displaystyle \bigodot$}\mskip-2.2mu \llap {$\ddot
        \smile$}}}

\def\Hat#1{\rlap{\kern.10em$\widehat{\phantom G}$}#1}
\def\HAt#1{\rlap{\kern.05em$\widehat{\phantom G}$}#1}

\def\cap#1{\rlap{\kern.1em$\widehat{\phantom{G\vrule height.8em}}$}#1{}}
\def\Cap#1{\rlap{\kern.05em$\widehat{\phantom{G\vrule height.8em}}$}#1{}}

\def\ket#1{\left| #1\right\rangle}

\def\VEV#1{\left\langle #1\right\rangle}

\def\abs#1{\left| #1\right|}
\def\leftrightarrowfill{$\mathsurround=0pt \mathord\leftarrow \mkern-6mu
        \cleaders\hbox{$\mkern-2mu \mathord- \mkern-2mu$}\hfill
        \mkern-6mu \mathord\rightarrow$}
\def\overleftrightarrow#1{\vbox{\ialign{##\crcr
        \leftrightarrowfill\crcr\noalign{\kern-1pt\nointerlineskip}
        $\hfil\displaystyle{#1}\hfil$\crcr}}}
 
\def\frac#1#2{{\textstyle{#1\over\vphantom2\smash{\raise.20ex
        \hbox{$\scriptstyle{#2}$}}}}}

\catcode`@=11
\def\underline#1{\relax\ifmmode\@@underline#1\else
        $\@@underline{\hbox{#1}}$\relax\fi}
\catcode`@=12
 
\def\nis{\nointerlineskip}
\def\Abar{\vbox{\nis\moveright.33em\vbox{
        \hrule width.35em height.04em}\nis\kern.05em\hbox{$A$}}{}}
\def\Dbar{\vbox{\nis\moveright.20em\vbox{
        \hrule width.50em height.04em}\nis\kern.05em\hbox{$D$}}{}}
\def\Gbar{\vbox{\nis\moveright.20em\vbox{
        \hrule width.50em height.04em}\nis\kern.05em\hbox{$G$}}{}}
\def\mbar{\vbox{\nis\moveright.15em\vbox{
        \hrule width.60em height.04em}\nis\kern.05em\hbox{$m$}}{}}
\def\Rbar{\vbox{\nis\moveright.20em\vbox{
        \hrule width.50em height.04em}\nis\kern.05em\hbox{$R$}}{}}
\def\Vbar{\vbox{\nis\moveright.05em\vbox{
        \hrule width.60em height.04em}\nis\kern.05em\hbox{$V$}}{}}
\def\Xbar{\vbox{\nis\moveright.20em\vbox{
        \hrule width.60em height.04em}\nis\kern.05em\hbox{$X$}}{}}
\def\thetabar{\vbox{\nis\moveright.15em\vbox{
        \hrule width.30em height.04em}\nis\kern.05em\hbox{$\theta$}}{}}
\def\Lambdabar{\vbox{\nis\moveright.25em\vbox{
        \hrule width.35em height.04em}\nis\kern.05em\hbox{${\mit\Lambda}$}}{}}
\def\Sigmabar{\vbox{\nis\moveright.25em\vbox{
        \hrule width.50em height.04em}\nis\kern.05em\hbox{${\mit\Sigma}$}}{}}
\def\phibar{\vbox{\nis\moveright.18em\vbox{
        \hrule width.40em height.04em}\nis\kern.05em\hbox{$\phi$}}{}}
\def\chibar{\vbox{\nis\moveright.12em\vbox{
        \hrule width.40em height.04em}\nis\kern.05em\hbox{$\chi$}}{}}
\def\psibar{\vbox{\nis\moveright.23em\vbox{
        \hrule width.40em height.04em}\nis\kern.05em\hbox{$\psi$}}{}}
\def\debar{\vbox{\nis\moveright.18em\vbox{
        \hrule width.35em height.04em}\nis\kern.05em\hbox{$\partial$}}{}}
\def\delbar{\vbox{\nis\moveright.10em\vbox{
        \hrule width.63em height.04em}\nis\kern.05em\hbox{$\nabla$}}{}}

\def\rarr{\rightarrow}

\begin{document}

\rightline{LA-UR-94-2479}
\rightline{August 24, 1994} \vskip .5in

\centerline{\large \bf  SEMIQUANTUM CHAOS AND } 
\centerline{\large \bf THE LARGE N EXPANSION}

\vspace{1cm}

\centerline{{\bf Fred Cooper$^{\star}$, John Dawson$^{\dagger}$, Salman
Habib$^{\star}$,}} 
\centerline{{\bf Yuval Kluger$^{\star}$, Dawn Meredith$^{\dagger}$,
and Harvey Shepard$^{\dagger}$}} 

\vspace{1cm}

\centerline{$^{\star}${\em Theoretical Division and Center for
Nonlinear Studies}} 
\centerline{{\em Los Alamos National Laboratory}}
\centerline{{\em Los Alamos, NM 87545}}

\vspace{.5cm}

\centerline{$^{\dagger}${\em Department of Physics}}
\centerline{{\em University of New Hampshire}}
\centerline{{\em Durham, NH 03824}}

\vspace{1cm}

\centerline {\bf Abstract}

\vspace{.5cm}

We consider the dynamical system consisting of a quantum degree of
freedom $A$ interacting with $N$ quantum oscillators described by the
Lagrangian   
\bq
L = {1\over 2}\dot{A}^2 + \sum_{i=1}^{N} \left\{
    {1\over 2}\dot{x}_i^2 - {1\over 2}( m^2 + e^2 A^2)
     x_i^2 \right\}~.
\eq
In the limit $N \rightarrow \infty$, with $e^2 N$ fixed, the quantum
fluctuations in $A$ are of order $1/ N$. In this limit, the $x$
oscillators behave as harmonic oscillators with a time dependent mass
determined by the solution of a semiclassical equation for the
expectation value $\VEV{A(t)}$. This system can be described, when
$\VEV{x(t)}= 0$, by a classical Hamiltonian for the variables $G(t) =
\VEV{x^2(t)}$, $\dot{G}(t)$, $A_c(t) = \VEV{A(t)}$, and
$\dot{A_c}(t)$. The dynamics of this latter system turns out to be
chaotic. We propose to study the nature of this large-$N$ limit by
considering both the exact quantum system as well as by studying an
expansion in powers of $1/N$ for the equations of motion using the
closed time path formalism of quantum dynamics.  

\newpage

\section{Introduction}

The definition and observation of chaotic behavior in classical
systems is familiar and more or less well understood
\cite{bib:classicalchaos}.  However the proper definition of chaos for
quantum systems and its experimental manifestations are still unclear
\cite{bib:quantumchaos}.  Here we first study a simple model of two
coupled systems which displays {\em semiquantum} chaos
\cite{bib:semichaos} when one of the systems can be treated
``semiclassically.'' We then study a purely quantum system of $N+1$
degrees of fredom which has the identical dynamics in the large $N$
limit as our original system. In this way we can determine, as a
function of $N$, what is the time scale for quantum fluctuations of
the ``classical'' oscillator to be of significant size. We also can
determine how this time scale is related to the time scale determined
by the maximum Lyapunov index. The question we are interested in here
is whether quantum fluctuations become significant before or after the
original system is sensitive to initial conditions.

In a classical chaotic system, such as the weather, we are accustomed
to situations where there is lack of long time forecasting because of
the sensitivity of the system to initial conditions. The simple model
system considered here has the unusual feature that one has to give up
long term forecasting even for the quantum mechanical probabilities,
as exemplified by the average number of quanta at later times
\cite{bib:semichaos}.

First let us review the original system which displayed semiquantum
chaos.  Consider two coupled quantum systems described by the
Lagrangian,
\bq
L = \half \dot{x}^2 + \half \dot{A}^2 
  - \half ( m^2 + e^2 A^2) x^2~.              \label{eq:Lag}
\eq
This Lagrangian leads to the Heisenberg equations of motion:
\ba
\ddot{x} + ( m^2 + e^2 A^2) x & = & 0  \label{eq:xdot} \\
\ddot{A} + e^2 x^2 \, A & = & 0~. \label{eq:Adot}
\ea
The Hamiltonian is
\bq
H = \half p^2 + \half \Pi_A^2 + \half ( m^2 + e^2 A^2) x^2~,
\label{eq:Hamilt} 
\eq
where $p(t) = \dot{x}(t)$ and $\Pi_A = \dot{A}(t)$. We next assume
that we are in an experimental situation where the expectation value
of $A(t)$ is so large that quantum fluctuations may be ignored. That
is, we assume that $A$ is in a classical domain or in a coherent state
(with large displacement). This is a particular assumption about
approximating the expectation values involved in taking expectation
values of the Heisenberg equations of motion, namely:
\bq
\VEV{A^2 x}= \VEV{A}^2 \VEV{x}~; ~~ \VEV{x^2 A} = \VEV{x^2} \VEV{A}
\eq

Here expectation value means taking a trace with respect to an initial
density matrix defined at the initial time $t_0$, which we take to be
$t_0=0$.  Taking expectation values of the Heisenberg equations with
the above factorization, we obtain
\ba
\VEV{\ddot{x}}  + ( m^2 + e^2 \VEV{A}^2)\VEV{x}& = & 0~,
\label{eq:xdot2} \\ 
\VEV{\ddot{A}}+ e^2 \VEV{x^2}\VEV{A} & = & 0~. \label{eq:Adot2}
\ea
In this approximation, the equation for $\VEV{x}$ is that of an
harmonic oscillator with a time dependent mass, $m^2(t) = m^2 +e^2
\VEV{A}^2$. $A$ can be thought of as a ``classical'' oscillator
(since we do not include its quantum fluctuations) whose mass is
determined by the quantum fluctuation of the $x$ oscillator (we
consider the case where $\VEV{x}$=0).  The problem of a quantum
harmonic oscillator with a time dependent mass can be solved in terms
of the (numerical) solution of an auxiliary classical oscillator
problem. We begin by noting that in the Heisenberg picture
\bq
[x(t), p(t)] = i~.                                 \label{eq:comm}
\eq
This commutation relation can be satisfied at all times by introducing
time-independent (defined at $t=0$) creation and destruction
operators, $a$ and $a^{\dagger}$, and using the Ansatz
\bq
x(t) = f(t)a + f^*(t)a^{\dagger}~,                     \label{eq:xfa}
\eq
with $f(t)$ satisfying the Wronskian condition
\bq   
i [f^*(t) \dot{f}(t) - \dot{f}^*(t) f(t)] = 1~.      \label{eq:Wron1}
\eq
The destruction and creation operators $a$ and $a^{\dagger}$ satisfy
the usual commutation relation $[a,a^{\dagger}] = 1$. The commutation
relation (\ref{eq:comm}) then follows automatically. 

It is easy to show using (\ref{eq:xdot}) and (\ref{eq:xfa}) that
$f(t)$ satisfies the equation of motion 
\bq
\ddot{f} + ( m^2 + e^2 A^2) f  =  0~,                 \label{eq:fdot}
\eq
with the normalization fixed by the Wronskian condition
(\ref{eq:Wron1}).  We can either solve this classical equation
directly numerically, imposing the Wronskian condition at time $t=0$,
or we can automatically impose the Wronskian condition by the
substitution
\bq
f(t) = {1\over\sqrt{2 \Omega(t)}}\exp\left[ - i\int_0^t
\Omega(t')dt'\right]~, 
\eq
where $\Omega(t)$ satisfies the nonlinear differential equation
\bq
\half\left( {\ddot{\Omega} \over \Omega} \right) - 
{3\over 4}\left( {\dot{\Omega} \over \Omega} \right)^2 + \Omega^2
    =  \omega^2~,                                     \label{eq:Omega}
\eq
with
\bq
\omega^2(t)  \equiv m^2 + e^2 A^2(t)~.                \label{eq:omega}
\eq
For simplicity we choose the initial state vector at $t=0$ to be the
ground state of the operator $\hat{n} = a^{\dagger}a$, {\em i.e.},
$|\Psi(0)\rangle=|0\rangle$, where $a|0\rangle=0$. Then, from
(\ref{eq:xfa}), the average (classical) value of $x(t)$ and $p(t)$ is
zero for all time, {\em i.e.}, $\VEV{x(t)}=0$ and $\VEV{p(t)}=0$. This
initial condition pertains in certain semiclassical time evolution
problems, such as particle production by strong electromagnetic or
gravitational fields. For the electric field problem, $A$ corresponds
to the electric field, and $x$ to the $k=0$ mode of the charged
particle field (See, {\em e.g.}, Refs.
\cite{bib:coopetal}\cite{ref:CHKMPA}). 
 
The quantum fluctuations of $x(t)$ are non-zero and are given
by the variable $G(t)$,
\bq
G(t) = \VEV{x^2(t)}=\abs{f(t)}^2 = {1 \over 2 \Omega(t)}~.
\eq
From ({\ref{eq:Omega}), it is easy to show that $G(t)$ satisfies
\bq
\half\left({\ddot{G}\over G}\right) -
{1\over 4}\left({\dot{G}\over G}\right)^2 - 
{1 \over 4 G^2} + \omega^2 = 0~.                      \label{eq:Gdd}
\eq
In addition, we find that
\bq
\VEV{\dot{x}^2(t)}= {1\over 4}\left({\dot{G}^2 \over G} + {1 \over
G}\right)~.
\eq
The expectation value of Eq. (\ref{eq:Hamilt}) becomes a new effective
Hamiltonian
\ba
H_{eff} &=& \VEV{H(t)}                               \nonumber \\
&=&\half\Pi_A^2 + 2 \Pi_G^2 G + 
        {1 \over 8 G} + \half ( m^2 + e^2 A^2 ) G~.   \label{eq:Heff}
\ea
The momenta conjugate to $G$ and $A$ are
\bq
\Pi_G = {\dot{G} \over 4 G}~,~~~~\Pi_A = \dot{A}~.     \label{eq:mom}
\eq
This {\em classical} Hamiltonian determines the variables, $G$ and
$\dot{G}$, necessary for a complete {\em quantum-mechanical}
description of the $x$ oscillator. Hamilton's equations then yield 
\ba
\dot{\Pi}_G  &=& - 2 \Pi_G^2 + {1\over 8 G^2 } - 
\half\omega^2~,                                           \nonumber \\
\dot{\Pi}_A  &=& - e^2 A G~,
\ea
or equivalently,
\begin{eqnarray}
\ddot{A} + e^2 G \, A &=& 0~,                             \nonumber \\
\half\left( {\ddot{G} \over G} \right) -
{1\over 4}\left({\dot{G} \over G}\right)^2 - 
{1 \over 4 G^2} + \omega^2  &=& 0~,                 \label{eq:classeq}
\ea
which correspond to (\ref{eq:Gdd}) and the expectation value of Eq.
(\ref{eq:Adot}).  

The classical effective Lagrangian corresponding to the effective
Hamiltonian (\ref{eq:Heff}) is
\bq
L_{eff}=\half{\dot{A}^2}+ {1\over 8}\left({\dot{G}^2 \over G} - 
        {1 \over G}\right) - \half( m^2 + e^2 A^2 ) G~. 
\eq
This Lagrangian could also have been obtained using Dirac's action,
\bq
\Gamma =\int dt \langle\Psi(t) | i {\pa\over \pa t} - H |\Psi(t)
\rangle \equiv \int dt \, L_{eff},                    \label{eq:Leff}
\eq
and a time-dependent Gaussian trial wave function as described in Ref.
\cite{bib:CPS}. This variational method has recently been used to 
study the quantum Henon-Heiles problem  in a mean-field approximation
\cite{bib:Patt}. In this method, the Gaussian trial wave function is
parametrized as follows 
\bq
\Psi(t)= [2 \pi G(t)]^{-1/4} \exp[ -(x-q(t))^2
( G^{-1}(t)/4 - i \Pi_G(t))+ ip(t)(x-q(t))] .
\eq
Here, $G(t)$ and $\Pi_G(t)$ are the time dependent real and imaginary
parts of the width of the wave function. One can prove for our problem
that if the  quantum oscillator wave function starts at $t=0$ as a
Gaussian, it is described at all times by the above expression, where
$G(t)$ and $\Pi_G(t)$ are totally determined by solving the effective
Hamiltonian dynamics. (For our special initial conditions
$p(t)=q(t)=0$). Thus we find that our effective Hamiltonian totally
determines the time evolution of the quantum oscillator.

One interesting ``classical'' variable is the expectation value of the
time dependent adiabatic number operator, which corresponds to the
number of quanta in a situation where the classical $A$ oscillator is
changing slowly (adiabatically). For the related field theory problem of
pair production of charged pairs by strong electric fields (where $A$
corresponds to the classical electromagnetic field and $x$ to the
$k=0$ mode of the charged scalar field) this corresponds to the time
dependent single particle distribution function of charged mesons. To
find the expression for the number of quanta, which requires the
definition of an adiabatic number operator, we begin with the wave
function for the quantum oscillators corresponding to a slowly varying
classical background $A$:
\bq
g(t) = {1\over \sqrt{2 \omega(t)}}\exp \left[-i\int_0^t \omega(t')
dt'\right]   
\eq
in terms of which we can decompose the quantum operator via
\bq
x(t) =  g(t)b(t) + g^{\ast}(t)b^{\dagger}(t)~.
\eq
Requiring  the momentum operator to have
the form 
\bq
p(t) = \dot{x}(t) =  \dot{g}(t)b(t) + \dot{g}^{\ast}(t)b^{\dagger}(t) 
\eq
by imposing $g(t)\dot{b}(t) + g^{\ast}(t)\dot{b}^{\dagger}(t) = 0 $,
and recognizing that $g(t)$ and $g^{\ast}(t)$  satisfy the
Wronskian condition by construction, one finds that $b(t)$ and
$b^{\dagger}(t)$ have the usual interpretation as creation and
annihilation operators, {\em i.e.}, $[x(t),p(t)] = i$ and $ [b(t),
b^{\dagger}(t)] = 1$. Note also that
\bq
b(t)  =  i [g^{\ast}(t)\dot{x}(t) - \dot{ g}^{\ast}(t)x(t) ]~.
\eq 
It turns out that $b^{\dagger}(t)b(t) $ can be interpreted as a {\em
time-dependent} number operator (assuming a slowly varying (adiabatic)
classical field $A$). The time independent basis and the time dependent
basis are both complete sets and are related by a unitary Bogoliubov
transformation, $b(t) = \alpha(t)a + \beta(t)a^{\dagger} $, where
\ba
\alpha(t) &=& i[g^{\ast}(t) \dot{f}(t) -  \dot{g}^{\ast}(t) f(t)]~,\\
\beta(t) &=& i[g^{\ast}(t)\dot{f}^{\ast}(t) -
\dot{g}^{\ast}(t)f^{\ast}(t)]~,   
\ea
and where $ |\alpha(t)|^2 - |\beta(t)|^2 = 1 $. If we choose as
initial conditions, $\Omega(0) = \omega(0)$, $\dot{\Omega}(0) =
\dot{\omega}(0)$, then $\alpha(0) = 1$ and $\beta(0) = 0$. These are
the initial conditions appropriate to the field theory problem of pair
production. The average value of the time-dependent occupation number
is given by 
\bq
n(t) = \VEV{b^{\dagger}(t) b(t)} = |\beta(t)|^2 = (4\Omega\omega )^{-1}
\left[ (\Omega - \omega)^2 + {1\over 4}\left({\dot{\Omega}\over\Omega} -
{\dot{\omega} \over \omega} \right)^2  \right]~.       \label{eq:aven}
\eq
Eq.~(\ref{eq:aven}) allows us to compute the average occupation 
number of the system as a function of time.  

\section{Numerical Solution of  Hamilton's equations}

We now summarize some previous results from the numerical solution of
the Hamiltonian equations obtained in the Gaussian approximation
\cite{bib:semichaos}. For calculational purposes, it turns out to be
convenient to scale out the mass via the transformations $t\rightarrow
m^{-1}t$, $A \rightarrow m^{-1/2}A$, $G\rightarrow m^{-1}G$, and
$e\rightarrow e m^{3/2}$. Then the scaled equations of motion are 
\ba
\ddot{A} + e^2 G \, A &=& 0 \nonumber \\
\half \left({\ddot{G} \over G} \right) -
{1\over 4}\left( {\dot{G} \over G} \right)^2 - 
   {1 \over 4 G^2} + 1 + e^2 A^2  & = & 0~. \label{eq:classeqscaled}
\ea
In order to explore the degree of chaos as a function of (scaled)
energy and coupling parameter $e$, we calculated surfaces of section
and Lyapunov exponents. The surface of section is a slice through the
three-dimensional energy shell \cite{bib:classicalchaos}. That is, for
a fixed energy and coupling parameter, the points on the surface of
section are generated as the trajectory pierces a fixed place ({\em
e.g.}, $A=0$) in a fixed direction. The hallmark of regular motion is
the cross section of a KAM torus which is seen as a closed curve in
the surface of section. The hallmark of chaotic motion is the lack of
any such pattern in the surface of section. In Fig. 1 we show a plot
of a surface of section at $E=0.8$ and $e=1$ where regular and chaotic
regions co-exist.

\begin{figure}
\centerline{\psfig{figure=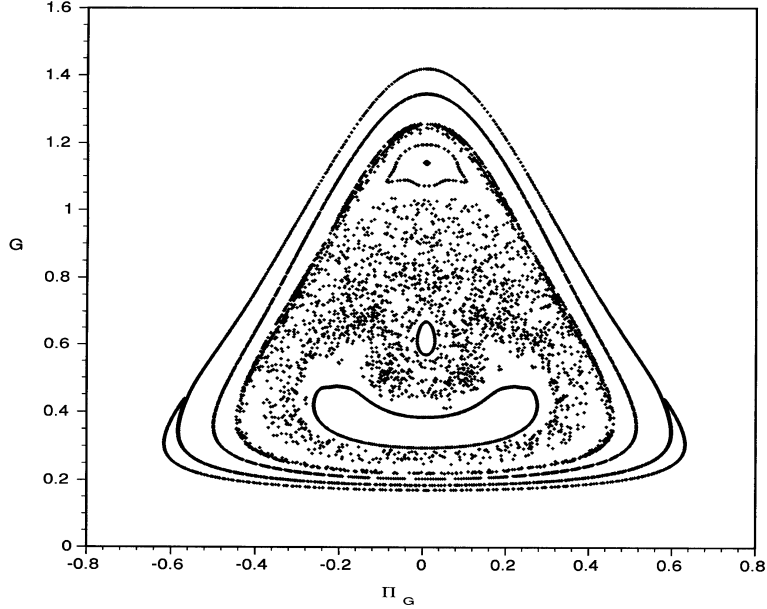,height=8cm,width=10cm}}
\caption[Figure 1] {\small{A plot of the surface of section for energy
= $0.8$, $e=1.0$, and $A=0$. The one chaotic region is in the center
of the plot.}} 
\end{figure}

The Lyapunov exponent provides a more quantitative, objective measure
of the degree of chaos. The Lyapunov exponent, $\lambda$, gives the
rate of exponential divergence of infinitesimally close trajectories
\cite{bib:lyapunov}. Although there are as many Lyapunov exponents as
degrees of freedom, it is common to simply give the largest of
these. For regular trajectories $\lambda =0$; for chaotic trajectories
the exponent is positive. To define the notion of a Lyapunov exponent
one begins by considering the infinitesimal deviation from a fiducial
trajectory, 
\bq
{\bf Z}(t) \equiv \lim_{\delta \rightarrow 0}
{\bf z} ({\bf z}_0 + \delta , t) - {\bf z} ({\bf z}_0,t)~,
\label{eq:eta} 
\eq
where ${\bf z} ({\bf z}_0 , t) $ is a point in phase space at time $t$
with initial position ${\bf z}_0$. The time evolution for ${\bf Z}(t)$
is given by 
\bq
{\dot {\bf Z}}(t)  = {\bf\nabla}{\bf F}\mid_{{\bf z}({\bf z}_0,
t)}\cdot {\bf Z}(t)~,               \label{eq:etatime}
\eq
where
\bq
{\dot {\bf z}} (t) = {\bf F} ({\bf z}(t),t)
\eq
are the full equations of motion for the system. The (largest)
Lyapunov exponent is defined to be  
\bq
\lambda \equiv \lim_{t \rightarrow \infty} { 1 \over t} \ln \left
\vert{{\bf Z}(t) \over {\bf Z}(0)}\right\vert~.    \label{eq:lyapunov}
\eq
Appendix {\em A} of Ref.~\cite{bib:lyapunov} provides an explicit
algorithm for the calculation of all the Lyapunov exponents. Since we
cannot carry out the $t \rightarrow \infty$ limit computationally, the
regular trajectories are those for which $\lambda(t)$ decreases as
$1/t$, while the chaotic trajectories give rise to $\lambda(t)$ that
is roughly constant in time, as judged by a linear least-squares fit
of $\log[\lambda(t)]$ {\em vs.}\ $\log(t)$.

We computed the Lyapunov exponents for three values of the scaled
coupling constant $e$ ($0.1$, $1.0$, $10.0$) and for energies from
$0.5$ to $2.0$. $E=0.5$ is the lowest energy possible, corresponding
to the zero point energy of the oscillator; there is no upper limit on
$E$. Fifty initial conditions were chosen at random for each energy
bin of width $0.1$ and coupling parameter. One relevant quantity to
study is the chaotic volume, the fraction of initial conditions with
positive definite Lyapunov exponents (corresponding to chaotic
behavior). Errors in this quantity arise because of the finite number
of initial conditions chosen, and because the distinction between zero
and positive exponents cannot be made with certainty at finite times.
We found that for $e=0.1$, more than $95\%$ of trajectories were
regular for all energies tested; for $e=1.0$ and $10.0$, there is a
steadily increasing fraction of chaotic orbits between $0.5 \le E \le
1.25$. For $1.25 \le E \le 2.0$, more than $90\%$ of these orbits are
chaotic. In Fig. 2 we show that the occupation number is sensitive to
initial conditions.

\begin{figure}
\centerline{\psfig{figure=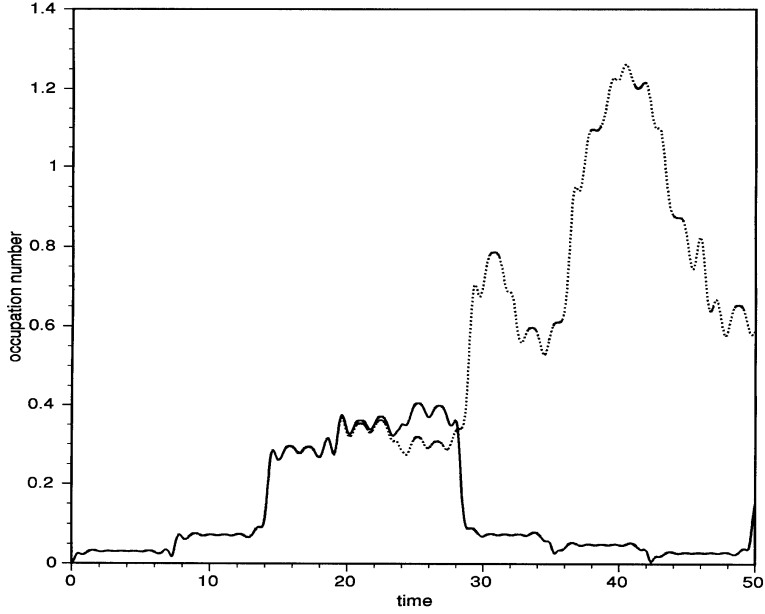,height=8cm,width=10cm}}
\caption[Figure 2] {\small{A plot of the occupation number given by
Eq.~(\ref{eq:aven}) for energy~=~ $1.8$, $e=1.0$, $A(0)=0$, $\Pi_G(0)=0$.  
The solid line is for for $G(0) = 0.5$; the dashed line is for $G(0)
=0.5001$. This plot shows the sensitivity to initial conditions.}} 
\end{figure}

\section{Quantum Oscillators and the Large-N Expansion}
\subsection{The Large-N Expansion and Semiquantum Chaos}

In this section we show that the previous system of equations for the
equations of motion for two oscillators are just the first term in a
large-$N$ expansion for expectation values of the operator equations
of motion of a quantum system consisting of $N$ copies of the original
$x$ oscillator ($ x \rightarrow x_i, i=1,2,...,N$) and a single quantum
dynamical variable $A$. Such a system is described by the operator
equations of motion:   
\ba
\ddot{x}_i + ( m^2 + e^2 A^2(t))x_i & = & 0~,            \nonumber \\
\ddot{A} + e^2 \left( \sum_{i=1} ^N x_i^2(t) \right) A & = &0~.
\ea
If all the $N$ quantum oscillators $x_i$ have the same initial
conditions then, at the level of expectation values, we can set all of
them equal ($\VEV{x_i}\equiv\VEV{x}$) and obtain equations of motion
for the expectation values:  
\ba
\VEV{\ddot{x}} + m^2 \VEV{x} + e^2 \VEV{A^2(t)x} &=& 0~,  \nonumber \\
\VEV{\ddot{A}} + e^2 N \VEV{x^2(t)  A} & = & 0~.
\ea

We are interested in initial conditions where $\VEV{x}=0$. In the
large N limit we will show that in this case 
\bq
\VEV{x^2 A} = \VEV{x^2}\VEV{A} + O(1/N)~;~~~~\VEV{A^2 x} = \VEV{A}^2
\VEV{x} + O(1/N), 
\eq
so that after the rescaling,
\bq
A \rightarrow \sqrt{N} \tilde{A}~;~~~~e^2 \rightarrow \tilde{e}^2/N~,
\eq
we recover the equations for the expectation values $\VEV{x^2}=G(t)$
and $\VEV{\tilde{A}}$ that pertained to semiquantum chaos. To compare
with the orginal system describing semiquantum chaos we must solve the
quantum system at different $N$ for fixed $\tilde{e}$. This value of
$\tilde{e}$ must then be set equal to the value of $e$ used in the
semiquantum chaos problem. We must also compare $\VEV{\tilde{A}(t)}$
to the classical oscillator motion $A(t)$ to see the effects of the
quantum fluctuations.

The exact quantum problem for the coupled quantum oscillators $x$ and
$A$ can be studied as a function of $N$ once we supply the initial
wave function at time $t=0$. Since we are interested in a comparison
with our previous calculation, where $\tilde{e}$ is kept fixed, it is
convenient to write the equations of motion as
\ba
\VEV{\ddot{x}} + m^2 \VEV{x} + \tilde{e}^2\VEV{{A^2(t)\over N}x} &=&
0~, \nonumber \\
\VEV{\ddot{A}} + \tilde{e}^2\VEV{x^2(t)  A}& = & 0~.
\ea
The initial conditions, as we change N, are that for fixed
$\tilde{A}_0(0)$ at time $t=0$,
\bq
\VEV{A(0)} = \sqrt{N} \tilde{A}_0(0)~;~~~~\VEV{x(0)} = 0~.
\eq

In the Schr\"odinger picture we have instead the time dependent
Schr\"odinger equation 
\bq
i{\partial \Psi(x_i,A,t)  \over \partial t} = \left[-{1\over 2}
\sum_i{\partial^2 \over \partial x_i^2} - {1\over 2} {\partial^2 \over
\partial A^2} + (m^2 + e^2 A^2) \sum_i x_i^2\right] \Psi (x_i,A,t)~. 
\eq
With an eye to convenience in solving the Schr\"odinger equation we
will choose an initial wave function which is a Gaussian in $A$  and
$x$, so that the initial expectation values of $A$ and $x$ will be the
ones specified above. This can be implemented by the initial wave
function: 
\bq
\Psi(t=0) =\prod_{i=1}^N \Psi_{x_i}(t=0) \Psi_A(t=0)~, \label{eq:wave}
\eq
\bq
\Psi_A(0) = [2 \pi D(0)]^{-1/4} \exp[ -(A-\sqrt{N}\tilde{A}_0)^2
(D^{-1}(0)/4 - i \Pi_D(0))+ ip_A(0)(A-\sqrt{N}\tilde{A}_0)]~.
\eq
with  
\bq 
D(0)+ N \tilde{A}_0^2 = \VEV{A^2}_{t=0}~,~~~~\Pi_D(0) = 
{\dot{D}(0)\over 4 D(0)}~,~~~~p_A(0) = \VEV{-i {\partial \over
\partial A}}_{t=0}~, 
\eq
and
\bq 
\Psi_{x_i}(0)= [2 \pi G(0)]^{-1/4} \exp\left[ -x_i^2
(G^{-1}(0)/4 - i \Pi_G(0))\right]~.
\eq
These initial conditions can be used to compare, as a function of $N$,
the exact quantum problem with the semiquantum chaos problem as well
as the $1/N$ correction to the semiquantum chaos problem. 

The large-$N$ expansion is best formulated using path integral methods
for the generating function of the expectation values. For initial
value problems rather special boundary conditions must be placed on
the Green's functions to insure causality. The formalism for doing
this is the the closed time path (CTP) formulation of the effective
action \cite{ref:SchKel}. The marrying of the large $N$ expansion to
the CTP formalism was accomplished recently, as described in Ref.
\cite{ref:CHKMPA}. 

Let us first ignore the issue of boundary conditions on the Green's
functions and discuss the generating functional for the expectation
values. To obtain the generating functional we add sources to the
original Lagrangian and consider the action 
\bq
S = \int_c ~ L~ dt
\eq
where the Lagrangian is,
\bq
L = \half\dot{A}^2 + \sum_{i=1}^{N} \left\{
     \half\dot{x}_i^2 - \half ( m^2 + e^2 A^2)
     x_i^2 + j_i x_i \right\} + J A~.        \label{forty}
\eq
The contour $c$ will be chosen in a way that enforces the correct
boundary conditions for taking expectation values of operators at an
initial time $t=0$. This will be discussed in the appendix. 

The generating functional for the expectation values is given by the
path integral: 
\bq
Z[J,j] = \int{\rm d}[A] \int {\rm d}[x_1] \ldots \int {\rm d}[x_N] \>
e^{ i S[A,x;J,j] }~.                                 \label{eq:ZSL}
\eq
Since (\ref{forty}) is  quadratic in the $x_i$ variables, we
may integrate over all of them in (\ref{eq:ZSL}) and obtain an
effective action, given by: 
\bq
Z[J,j] = \int {\rm d}[A]e^{ i S_{eff}[A;J,j]}
\eq
where
\ba
S_{eff}[A;J,j] & = &  \int_{c} dt \left\{ - \half A { {\rm d}^2 A
\over  {\rm d} t^2 } + J A  \right\} + {i N \over 2} {\rm Tr~ \ln } [
G^{-1}(A) ] \nonumber \\  
& &{}+ \half \int_{c} dt \int_{c} dt' \sum_i \left\{ j_i(t) G(t,t')
j_i(t')  \right\}~.                  \label{eq:ZeffS}
\ea
Here, we have defined
\bq
G_0^{-1}(t,t';A)= \left\{{ {\rm d}^2\over{\rm d} t^2 } + ( m^2 + e^2
A^2 ) \right\}\delta(t - t')~.
\eq
Thus the Green's function $G_0(t,t')$ obeys
\bq
\left\{{{\rm d}^2\over{\rm d}t^2} + ( m^2 + e^2 A^2 )\right\}
G_0(t,t'')  =  \delta(t - t'')~.
\eq
The boundary conditions on this Green's function needed to insure
causality will be discussed later. We now consider the particular
situation with all the  $x_i$ identical ({\em i.e.}, have identical
initial conditions) so that $j_i = j,~x_i=x$. We also rescale $A$, $J$
and $e$ as follows: 
\ba
\tilde{A} & = & A/\sqrt{N} \nonumber \\
\tilde{J} & = & J/\sqrt{N} \nonumber \\
\tilde{e} & = & e \sqrt{N}~.              \label{eq:scaling}
\ea
The effective interaction now becomes proportional to $N$ as long as
$\tilde{e}^2$ is kept fixed when $N$ is changed. This allows the
evaluation of the remaining path integral over $A$ by the method of
steepest descent and leads to an expansion of the expectation values
as a power series in $1/N$. The value of the path integral at the
stationary point is the leading term in this expansion, and the
Gaussian fluctuations about the stationary point give the $1/N$
correction. Expanding the effective action about the stationary point:  
\ba
S_{eff}[A;J,j] &=& S_{eff}[A_0;J,j]        \nonumber \\ 
&& + \int_{c} dt' \left[ { \delta S_{eff} \over \delta A(t') }
\right]_{A_0} \left( A(t') - A_0(t') \right)  \nonumber \\ 
&& + \int_{c} dt' \int_{c} dt'' \left[{ \delta^2 S_{eff}\over \delta
A(t') \delta A(t'')}\right]_{A_0}\left( A(t') - A_0(t') \right)\left(
A(t'') - A_0(t'') \right)                     \nonumber \\ 
&& + \cdots 
\ea
The field $A_0$ is determined by the requirement
\bq
\left[{\delta S_{eff} \over \delta A(t) } \right]_{A_0} = 0~,
\eq
and setting,
\bq
D^{-1}_0(t,t') = - \left[{ \delta^2 S_{eff} \over \delta A(t) \delta
A(t')} \right]_{A_0}~, 
\eq
the path integral in Eq.~(\ref{eq:ZeffS}), including terms up to 
$1/N$, is given by
\ba
Z[J,j] & = & e^{i W[J,j]}                     \nonumber \\
W[J,j] & = & S_{eff}[A_0;J,j] + {i\over 2}{\rm Tr}\ln [
D^{-1}_0(x_0,A_0) ]  + \cdots~.                 \label{eq:ZW}
\ea
Since the first term in the action is proportional to $N$ and the
Gaussian fluctuation contribution is of order $N^0$, the fluctuation
term gives the $1/N$ corrections. An auxiliary quantity which allows
the direct determination of one particle irreducible vertices such as
inverse Green's functions is the effective action functional $\Gamma$
(not to be confused with $S_{eff}$) which is a Legendre transform of
$W[J,j]$. Changing variables from $J,j$ to the expectation values
$\VEV{x}$, $\VEV{A}$ where
\bq
\VEV{x} \equiv \bar{x} = {\delta W \over \delta j}~;
\hspace{.2in}\VEV{A} \equiv \bar{A}= {\delta W \over \delta J}~.
\eq
We now define the effective action functional (omitting the overline
for symplicity of notation) as
\bq
\Gamma[x_i,A] = W[j,J] -\int_c ~dt ~[j_i(t) x_i(t) + J(t)A(t)]~,
\eq
which turns out to be, at order $1/N$,
\bq
\Gamma[x_i,A] =S_{cl}[x_i,A] + {iN \over 2} {\rm Tr} \,  \ln [
G^{-1}_0(A) ] + {i\over 2}{\rm Tr} \,  \ln [ D^{-1}_0(x,A) ] 
\eq
where $S_{cl}$ is the classical action and where $A$ and $x$ are the
expectation values of the Heisenberg operators accurate to order $1/N$. 

We now demonstrate that at the stationary phase point the original
problem is recovered. When the sources are set to zero,
\bq
\left[ { \delta S_{eff} \over \delta A(t) } \right]_{A_0} = 
- \left\{ { {\rm d}^2  \over {\rm d}t^2 } + e^2 \left[\sum_ix_{0
i}^2(t) + {N \over i} G_0(t,t;A_0) \right] \right\} A_0(t) \equiv 0~,
\label{eq:derivi}
\eq
or after rescaling:
\bq
\left\{ { {\rm d}^2  \over {\rm d}t^2 } + \tilde{e}^2 \left[x_0^2(t) +
{1\over i} G_0(t,t;A_0) \right] \right\} \tilde{A}_0(t) = 0~,
\label{eq:A0} 
\eq
where $G_0(t,t';A_0)$ satisfies
\bq
\left\{ { {\rm d}^2 \over  {\rm d} t^2 } +  ( m^2 + e^2 A_0^2 )
\right\}  G_0(t,t';A_0)  = \delta(t - t')~,
\eq
and $x_{0i}(t)$ satisfies
\bq
\left\{ { {\rm d}^2  \over {\rm d} t^2 } +  ( m^2 + e^2 A_0^2 )
\right\} x_{0i }(t)  = 0~.                          \label{eq:xj0}
\eq
With the identification
\bq 
{1\over i} G_0(t,t) = G(t)
\eq
where $G(t) = \VEV{x^2(t)} - \VEV{x(t)}^2$, we arrive at the equations
we studied earlier pertaining to semiquantum chaos. (For that problem
we chose $\VEV{x(t)} = 0$.) Thus we have shown that the lowest
order in $1/N$ solution to the problem displays semiquantum chaos
(note that the quantity $e^2 A^2$ is invariant under our rescaling). 
 
In the above, the inverse propagator for the $A$ variable is given by:
\bq 
D^{-1}_0(t,t';A_0) = - \left[{ \delta^2 S_{eff} \over \delta A(t)
\delta A(t')} \right]_{A_0} = d^{-1}_0(t,t';A_0) +
\Pi^{\vphantom{-1}}_0(t,t';A_0)~,                \label{eq:Dinverse}
\eq
where
\ba
d^{-1}_0(t,t';A_0) & = & \left\{ { {\rm d}^2  \over  {\rm d}t^2 } 
+ e^2 \left[ \sum_i x_{0 i}^2(t) + {N \over i} G_0(t,t;A_0)
\right]\right\} \delta(t - t')~,                \nonumber \\
\Pi_0(t,t';A_0) & = & 2N e^4 A_0(t) \pi_0(t,t';A_0) A_0(t')~,
\nonumber \\ 
\pi_0(t,t';A_0) & = & i \, G_0(t,t';A_0) \, G_0(t',t;A_0) - {2 \over
N}  \sum_ix_{0 i}(t) G_0(t,t';A_0) x_{0 i}(t')~. \label{eq:dPipi}
\ea
Rescaling we have:
\ba
d^{-1}_0(t,t';A_0) & = & \left\{ { {\rm d}^2  \over  {\rm d}t^2 } +
\tilde{e}^2 \left[ x^2(t) + {1\over i} G_0(t,t;A_0) \right]\right\}
\delta(t - t')~,                                      \nonumber \\
\Pi_0(t,t';A_0) & = & 2\tilde{e}^4  \tilde{A}_0(t) \pi_0(t,t';A_0)
\tilde{A}_0(t')~,                                    \nonumber \\
\pi_0(t,t';A_0) & = & i \, G_0(t,t';A_0) G_0(t',t;A_0) - 2   x(t)
G_0(t,t';A_0)x(t')~.                           \label{eq:dPipi2}
\ea
We need to invert (\ref{eq:Dinverse}) subject to the correct causal
boundary conditions to find $D_0(t_1,t_2;A_0)$.

The causal Green's function for initial value problems can be
expressed in two ways, either as a two dimensional matrix Green's
function, or as a path ordered Green's function defined on a complex
contour. In this paper we will use the second method.  As discussed in
the appendix, the quantity that takes the place of the usual Feynman
Green's function is the causal Green's function. We begin by defining
an initial density matrix at time $t=0$ by $\rho$, and then
introducing the two Wightman functions $G_>$ and $G_<$ via 
\ba
G_{>}(t,t')&=& i \{\VEV{x(t) x(t')} - \VEV{x(t)}\VEV{x(t')}\} \nonumber \\
G_<(t,t')&=& i \{\VEV{x(t') x(t)} - \VEV{x(t')}\VEV{x(t)}\}
\ea
where $\VEV{x(t) x(t')} \equiv Tr\{\rho x(t) x(t')\}$. Time integrals
are then defined on the contour shown in Fig. 3, with the integration
path given by 
\bq
\int_{c} dt = \int_{0: c_+}^{\infty} dt - \int_{0: c_-}^{\infty} dt~.
\eq
\begin{figure}
\centerline{\psfig{figure=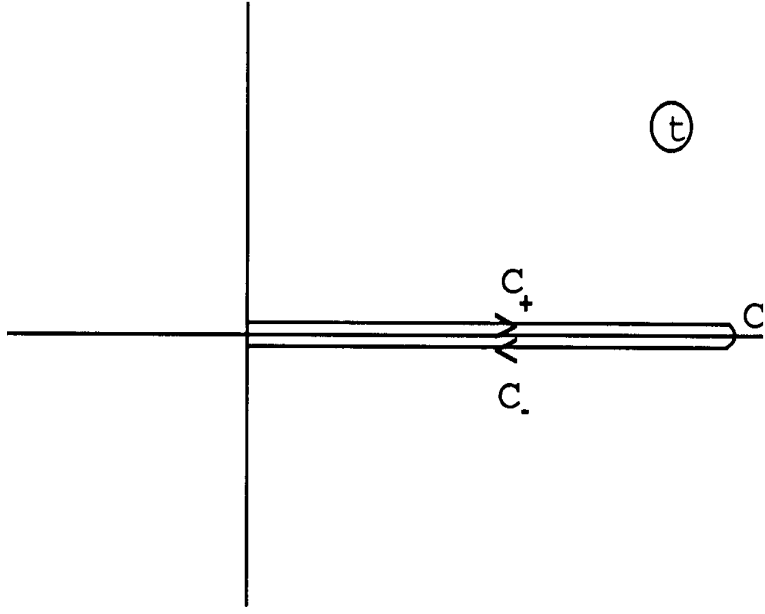,height=8.0cm,width=10.0cm}}
\caption[Figure 3]{Complex time contour $\cc$ for the closed time path
propagators.}
\end{figure} 
The causal Green's functions which embody the correct boundary
conditions are then 
\bq
G(t,t') = \Theta_{c}(t,t') G_{>}(t,t') + \Theta_{c}(t',t) G_{<}(t,t')~,
\eq
where
\begin{equation}
\Theta_{c}(t,t') = \cases{ 
\Theta(t,t') & for $t$ on $c_+$ and $t'$ on $c_+$ \cr 
0            & for $t$ on $c_+$ and $t'$ on $c_-$ \cr
1            & for $t$ on $c_-$ and $t'$ on $c_+$ \cr
\Theta(t',t) & for $t$ on $c_-$ and $t'$ on $c_-$ }~. \label{eq:theta}
\eq
The Green's functions are symmetric and $ A_{>}(t,t') = A_{<}(t',t) $.
These propagators take the place of the usual Feynman ones for initial
value problems. To calculate the $1/N$ corrections to the quantities
$\VEV{A(t)}$ and $\VEV{x(t)}$ and $\VEV{x^2(t)}$ one needs to take
functional derivatives of the generating functional, all correct to
order $1/N$. Specifically we have:
\bq
\VEV{A(t)} = {\delta W[J,j] \over \delta J(t)}
\eq
where $W[J,j]$ is given in Eq.~(\ref{eq:ZW}).  

To separate out the leading and next to leading order terms, we write,
\bq
\VEV{A(t)} =   A_0(t) +  A_1(t)~,
\eq
where $A_0(t)$ is the quantity determined in first order, by
Eq.~(\ref{eq:A0}).  $A_1(t)$ is given by 
\bq
A_1(t) = {i\over 2} \int_{c}dt'\int_{c}dt_1 \int_{c}dt_2 D_0(t,t';A_0) 
\left( { \delta D_0^{-1}(t_1,t_2;A_0)  \over  \delta A_0(t') } 
\right) D_0(t_2,t_1;A_0)~.                             \label{eq:a1}
\eq
By straightforward differentiation, it is possible to show that
$A_1(t)$ gets contributions from three terms  when $\VEV{x} = 0$.
Namely, 
\bq
A_1(t) = A_1^{(a)}(t) + A_1^{(b)}(t) + A_1^{(c)}(t)~,
\eq
where
\ba
A_1^{(a)}(t) & = & { i \tilde{e}^4 \over N} \int_{c}dt_1
\int_{c} dt_2 K^{(a)}(t,t_2,t_3)~,               \nonumber \\
K^{(a)}(t,t_2,t_3) & = & D_0(t,t_2;A_0) A_0(t_2) \pi_0(t_2,t_3;A_0)
D_0(t_3,t_3;A_0)~,                                         \\
A_1^{(b)}(t) & = & { 2 i \tilde{e}^4 \over N}\int_{c}dt_2
\int_{c}dt_3 K^{(b)}(t,t_2,t_3)~,                \nonumber \\
K^{(b)}(t,t_2,t_3) & = & D_0(t,t_2;A_0) \pi_0(t_2,t_3;A_0)
D_0(t_2,t_3;A_0) A_0(t_3)~,                                \\
A_1^{(c)}(t) & = & -{ 4 i \tilde{e}^6 \over N} \int_{c} dt_2
\int_{c} dt_3 \int_{c} dt_4 K^{(c)}(t,t_2,t_3,t_4)~,\nonumber \\
K^{(c)}(t,t_2,t_3,t_4) & = & D_0(t,t_2;A_0) \tilde{A}_0(t_2)
G_0(t_2,t_3;A_0) \tilde{A}_0(t_3)                 \nonumber \\
&&\times G_0(t_2,t_4;A_0) A_0(t_4) \Sigma (t_3,t_4;A_0)~.
\ea
In the above expressions, the self energy $\S$ is given by
\bq
\Sigma (t_3,t_4;A_0)= i D_0(t_3,t_4;A_0) G_0(t_3,t_4;A_0)~,
\eq
and the polarization $\pi$ by
\bq
\pi(t_3,t_4;A_0)= iG_0(t_3,t_4;A_0)G_0(t_3,t_4;A_0)~.
\eq
Note that the causal Green's functions are symmetric under the
interchange of time labels. It is easy to show, using the rules for
convoluting the causal Green's functions (given below), that $A_1(t)$
only depends on information from previous times. 

In order to evaluate the above graphs we first have to determine the
causal propagator $D_0$. From (\ref{eq:Dinverse}), we have
\bq
D^{-1}_0(t,t') = d^{-1}_0(t,t') + \Pi_0(t,t')~,
\eq
where $d^{-1}_0(t,t')$ and $\Pi_0(t,t')$ are defined in
(\ref{eq:dPipi}). The defining equations for the inverses are 
\ba
\int_{c} dt' \, d^{-1}_0(t,t') d^{\vphantom{-1}}_0(t',t'') & = &
\delta(t - t'')~,                                         \nonumber \\ 
\int_{c} dt' \, D^{-1}_0(t,t') D^{\vphantom{-1}}_0(t',t'') & = &
\delta(t - t'')~.                                \label{eq:dDinv}
\ea
If we now put
\bq
D_0(t,t') = d_0(t,t') + \tilde{D}_0(t,t')~,
\eq
then (\ref{eq:dDinv}) implies
\bq
\int_{c} dt'\left\{d^{-1}_0(t,t') + \Pi^{\vphantom{-1}}_0(t,t')
\right\} D^{\vphantom{-1}}_0(t',t'') = \delta(t - t'')~,
\eq
or
\bq
\int_{c} dt' d^{-1}_0(t,t') \tilde{D}^{\vphantom{-1}}_0(t,t')  = -
\int_{c} {\rm d}t' \, \Pi^{\vphantom{-1}}_0(t,t')
D^{\vphantom{-1}}_0(t',t'')~. 
\eq
We can now invert the left hand side using (\ref{eq:dDinv}) again,
obtaining: 
\bq
D_0(t_1,t_4) = d_0(t_1,t_4) - \int_{c} dt_2 \int_{c} dt_3 \,
d_0(t_1,t_2) \Pi_0(t_2,t_3) D_0(t_3,t_4)~.           \label{eq:DPi}
\eq
We use (\ref{eq:DPi}) to find the inverse. Note that $d_0(t,t')$
satisfies 
\bq
\left\{ { {\rm d}^2  \over  {\rm d}t^2 } + e^2 \left[\sum_i x_{0
i}^2(t) + {N \over i} G_0(t,t) \right] \right\} \, d_0(t,t') =
\delta(t - t')~. 
\eq
We will need to solve for this Green's function $d_0$ in order to find
$D_0(t,t')$ from (\ref{eq:DPi}). As shown below this can be done in
analogy to the determination of $G(t)$ in the semiquantum problem,
that is, we will introduce a set of mode functions for an auxiliary
quantum problem. 

To study the time it takes for the $1/N$ corrections to become
significant, we also need to determine the $1/N$ correction to the $x$
oscillator propagator.  This is obtained once again by functional
differentiation. Now we turn to the full Green's function
$\bar{G}(t,t')$  for the $x$ oscillator (to order $1/N$), from which
we can obtain 
\bq
\VEV{x^2(t)} = {1 \over i} \bar{G} (t,t).
\eq
This is most easily determined to order $1/N$ from the inverse
Green's function which by the chain rule is the negative of the second
derivative of the effective action functional $\Gamma[x,A]$ with
respect to $x$. The effective action functional $\Gamma$ to order
$1/N$ is 
\bq
\Gamma[x_i,A] =S_{cl}[x_i,A] + i {N \over 2} {\rm Tr} \ln [
G^{-1}_0(A) ] + { i \over 2} {\rm Tr} \ln [ D^{-1}_0(x,A) ]~, 
\eq
where $S_{cl}$ is the classical action and here $A$ and $x$ are the 
expectation values of the Heisenberg operators accurate to order
$1/N$, {\em i.e.}, the $A$ in this equation is the solution of $\delta
\Gamma /\delta A = 0$, or
\bq
\left\{ { {\rm d}^2  \over  {\rm d}t^2 } + e^2 \left[
\sum_i x_{0 i}^2(t) + {N \over i} G_0(t,t;A) \right] 
\right\} \, A(t) = {i \over 2} \int_c ~dt_1 \int_c ~dt_2
D_0 (t_1,t_2;A) {\delta D_0(t_2,t_1;A)  \over \delta A(t)}  \\
\eq
and differs from the stationary point $A_0$  of the original path integral
by terms of order $1/N$. Now
\ba
\bar{G}^{-1}_{ij}  (t,t';A)& = & -{\delta^2 \Gamma \over \delta x_i(t)
\delta x_j(t')}                                    \nonumber \\
&=& G_0 ^{-1}(t,t';A) \delta_{ij} -{i \over 2} \int_c ~dt_1 \int_c
~dt_2 D_0 (t_1,t_2;A) {\delta^2 D_0(t_2,t_1;A)  \over \delta x_i(t)
\delta x_j(t')}                                    \nonumber \\
&=& \delta_{ij} \bar{G}^{-1}(t,t';A)~.
\ea
And, since
\ba
{\delta^2 D_0(t_2,t_1;A)  \over \delta x_i(t) \delta x_j(t')}&=& 2 e^2
\delta(t_1-t_2) \delta (t_1-t) \delta(t_2-t')      \nonumber \\ 
&-& 4 e^4 A(t) G_0(t,t';A) A(t')[\delta (t_1-t) \delta(t_2-t')+\delta
(t_1-t') \delta(t_2-t)]~,                          \nonumber \\
\ea
we obtain
\bq
\bar{G}^{-1}(t,t';A) = {G_0}^{-1}(t,t';A)+\bar{G_1}^{-1}(t,t;A),
\eq
where $\bar{G_1}^{-1}$ is of order $1/N$ and given by two terms.
Explicitly, 
\bq
\bar{G_1}^{-1}(t,t') = -i {\tilde{e}^2 \over N} D_0(t,t) \delta (t-t')
+ 4 {\tilde{e}^4 \over N}  \tilde{A}_0(t) \Sigma(t,t')
\tilde{A}_0(t')~.  
\eq
To obtain the $1/N$ expansion of $\bar{G}$ we  have to invert this to
order $1/N$. We first need to reexpand $G_0^{-1}(A)$ up to order $1/N$
since $A=A_0+ A_1$ where $A_1$ is given by (\ref{eq:a1}). We have,
\ba
\bar{G}^{-1}(t,t') &=& {G_0}^{-1}(A_0)(t,t')+2 e^2 A_0(t) A_1(t)
\delta(t-t') +\bar{G_1}^{-1}(t,t')                  \nonumber \\
&\equiv& {G_0}^{-1}(A_0)(t,t')+ {1 \over N} \Delta (t,t')~.
\ea
Inverting to order $1/N$ we obtain finally
\bq
\bar{G}(t,t') = G_0(t,t';A_0) - {1 \over N} \int_c dt_1~ \int_c dt_2~
G_0(t,t_1;A_0) \Delta (t_1,t_2) G_0(t_2,t';A_0)+O(1/N^2)~.   \\
\eq

\subsection{Closed time path contour and causality}

In this section we discuss the causality of the Green's functions as
given by the CTP formalism. We will evaluate the time integrals using
the closed time path (CTP) contour shown in Fig. 3. The integration
path is given explicitly by
\bq
\int_{c} dt = \int_{0: c_+}^{\infty} dt - \int_{0: c_-}^{\infty} dt~.
\eq
The Green's functions are now given by functions of the form,
\bq
A(t,t') = \Theta_{c}(t,t') A_{>}(t,t') + \Theta_{c}(t',t) A_{<}(t,t')~,
\eq
where $\Theta_{c}(t,t')$ is defined in  Eq. (\ref{eq:theta}). We call
such functions {\em causal}.  The causal Green's functions are
symmetric implying thereby that $ A_{>}(t,t') = A_{<}(t',t)$. In order
to prove causality of any particular graph we need to discuss two lemmas.
The first is that if we have a loop of two causal functions, such as
found in a self energy graph, then that is also a causal function. To
show this one just needs the definition of $\Theta_c$. If the two
causal functions are
\ba
B(t,t') & = & \Theta_{c}(t,t') B_{>}(t,t') + \Theta_{c}(t',t)
B_{<}(t,t')~,                                          \nonumber \\
C(t,t') & = & \Theta_{c}(t,t') C_{>}(t,t') + \Theta_{c}(t',t)
C_{<}(t,t')~, 
\ea
then the self energy graph is also causal. Letting the self energy
\bq
A(t,t') = i B(t,t') C(t,t')~,
\eq
and setting,
\bq
A_{>,<}(t,t')= i B_{>,<}(t,t')C_{>,<}(t,t')~,
\eq
we find
\bq
A(t,t') = \Theta_{c}(t,t') A_{>}(t,t') + \Theta_{c}(t',t) A_{<}(t,t')~,
                                                  \label{eq:lemma} 
\eq
which is the desired result.

The next lemma is that the matrix product of two causal functions is
causal. That is, if we think of the Green's functions as matrices in
time and if we matrix multiply B and C so that  $A(t_1,t_3)$ is given
by 
\bq
A(t_1,t_3) =  \int_{c} {\rm d}t_2 \, B(t_1,t_2) C(t_2,t_3)~,
\eq
we find then
\bq
A(t,t') = \Theta_{c}(t,t') A_{>}(t,t') + \Theta_{c}(t',t) A_{<}(t,t')~,
                                                  \label{eq:lemma2}
\eq
where
\ba
A_{ \stackrel{>}{<} }(t_1,t_3) & = & - \int_{0}^{t_3} {\rm d}t_2 
B_{ \stackrel{>}{<} }(t_1,t_2) \left[ C_{>}(t_2,t_3) - C_{<}(t_2,t_3)
\right]                                             \nonumber \\
&&+ \int_{0}^{t_1} {\rm d}t_2 \left[ B_{>}(t_1,t_2) - B_{<}(t_1,t_2)
\right] C_{ \stackrel{>}{<} }(t_2,t_3)~.        \label{eq:matrixmult}
\ea
This lemma is discussed in Ref. \cite{ref:CHKMPA} and is obtained
directly by breaking the time integration into three segments, {\em
viz.}, 
\ba &(i)&  0 < t_i < t, \ {\rm on} \ \cc_+~, \nonumber \\ 
&(ii)&  t < t_i < \infty,\ {\rm on}\ \cc_+~,\nonumber \\ 
&(iii)&  0 < t_i < \infty,\ {\rm on}\ \cc_-~. 
\ea 
One then uses the definition of the $\Theta$ function (\ref{eq:theta})
and collects all the non-cancelling terms.

Now consider the product of three causal functions:
\bq
A(t_1,t_4)=\int_{c}dt_2 \int_{c}dt_3 B(t_1,t_2) C(t_2,t_3) D(t_3,t_4)~.
\eq
We can work this case out by applying the second lemma from left to
right. That is, we can let
\bq
E(t_1,t_3) = \int_{c} dt_2 \, B(t_1,t_2) C(t_2,t_3)~.
\eq
Then $E(t_1,t_3)$ is causal and is given by an equation of the form
(\ref{eq:lemma2}).  We are then left with an equation of the form:
\bq
A(t_1,t_4) =  \int_{c} dt_3 E(t_1,t_3) D(t_3,t_4)~,
\eq
which is also causal. In the same way, we can find causal relations
for any number of CTP integrals of causal functions. We can also
apply the lemma from right to left. After doing the integrals
sequentially one is eventually left with
\bq
f(t) = \int_c dt_1 F(t,t_1) = \int_0^{t} \left[F_{>}(t,t_1)
-F_{<}(t,t_1)\right]~, 
\eq
which explicitly displays the causality. 

\subsection{Lowest order causal Green's functions}

With the results of the previous section in hand we are now in a
position to solve for the Green's functions with the correct causal
structure. We would like to solve 
\ba
\left\{ { {\rm d}^2  \over {\rm d} t^2 } + ( m^2 + \tilde{ e}^2
\tilde{A}_0^2(t) ) \right\} G_0(t,t') & = & \delta(t - t')~,\nonumber
\\ 
\left\{ { {\rm d}^2 \over {\rm d}t^2 } + \tilde{e}^2 \left[x_{0 }^2(t)
+{1\over i}G_0(t,t)\right] \right\} d_0(t,t') & = & \delta(t - t')~, 
\label{eq:Greens}
\ea
subject to causal boundary conditions and an initial density matrix which
is that of an adiabatic vacuum at time zero. The adiabatic requirement
is satisfied by considering  auxiliary quantum fluctuation operators
$x_0$ and $A_{0q}$ such that $\VEV{x_0} = \VEV{A_{0q}} = 0$. (Note
that we are defining the fluctuation operators by writing the
Heisenberg operator $A=\VEV{A}+A_q$.) These operators are defined via
\ba
x_{0 }^{\vphantom{\ast}}(t)&=& f(t)a + f^{\ast}(t)a^{\ast} \nonumber \\
A_{0q}(t)   & = & g(t) b + g^{\ast}(t) b^{\ast}~,      \label{eq:xAab}
\ea
where $a$ and $b$ are canonical annihilation (creation) operators
satisfying 
\bq
[a,a^{\dagger}] = [b,b^{\dagger}] =1~, 
\eq
and the adiabatic vacuum is defined by
\bq
a  \ket{0}_{ad} = b \ket{0}_{ad} = 0~.
\eq
The $f(t)$ and $g(t)$ are functions of time satisfying  the
homogeneous equations, 
\ba
\left\{{{\rm d}^2\over{\rm d} t^2} + (m^2 + \tilde{ e}^2
\tilde{A}_0^2(t)) \right\}  f(t)  & = & 0                \nonumber \\
\left\{ { {\rm d}^2 \over {\rm d}t^2 } +\tilde{e}^2 \left[x_{0 }^2(t)
+ {1 \over i} G_0(t,t) \right]\right\} g(t) & = & 0~,  \label{eq:fgeqn}
\ea
with Wronskian conditions,
\ba
i \biggl\{ f^{\ast}(t)\overleftrightarrow{{{\rm d} \over {\rm d}t}}
f(t) \biggr\} & = & 1                                    \nonumber \\ 
i \biggl\{ g^{\ast}(t)\overleftrightarrow{{{\rm d} \over {\rm d}t}}
g(t) \biggr\} & = & 1~.                                \label{eq:Wron}
\ea
We can write the causal Green's functions in terms of the complex
functions $f(t)$ and $g(t)$,as follows:
\ba
G_0(t,t') & = & \Theta_{c}(t,t') G_{0>}(t,t')+\Theta_{c}(t',t) G_{0<}
(t,t')                                               \nonumber \\
d_0(t,t') & = &  \Theta_{c}(t,t') d_{>}(t,t') + \Theta_{c}(t',t)
d_{<}(t,t')~, 
\ea
where
\ba
G_{0>}(t,t') &= &\VEV{x_0(t) x_0(t')} =  i f(t) f^{\ast} (t')~, \\
G_{0<}(t,t') &= &\VEV{x_0(t') x_0(t)} =  i f(t') f^{\ast} (t)~,  \\
d_>(t,t') &= &\VEV{A_{0q}(t) A_{0q}(t')} =  ig(t)g^{\ast} (t')~, \\
d_<(t,t') &= &\VEV{A_{0q}(t') A_{0q}(t)} =  ig(t')g^{\ast} (t)~.
\ea
Note also that 
\ba
G_0(t,t) & = & i | f(t) |^2 = i G(t)~,                   \nonumber \\ 
d_0(t,t) & = & i | g(t) |^2~.
\ea
Here the factors of $i$ are introduced to agree with earlier
definitions. We are now in a position of being able to find the $1/N$
quantum corrections to $A_1(t)$ and $\VEV{x^2(t)}$, using Eq.
(\ref{eq:DPi}) to construct $D_0(t,t')$. The last-named quantity obeys
the integral equation 
\bq
D_0(t_1,t_4) = d_0(t_1,t_4) - \int_{c} dt_2 \int_{c} dt_3 d_0(t_1,t_2)
\Pi_0(t_2,t_3) D_0(t_3,t_4)~.  
\eq 
We want to evaluate this in terms of the lowest order quantities. The
polarization can be put in causal form since 
\bq
\pi_0(t,t') = i [ \Theta_{c}(t,t') G_{0>}^2 (t,t')+\Theta_{c}(t',t) 
G_{0<}^2 (t,t')]
\eq
and
\ba
\Pi_0(t,t') &=& 2 \tilde{e}^4 \tilde{A}_0(t)
\pi_0(t,t')\tilde{A}_0(t')~. 
\nonumber \\
&\equiv& \Theta_{c}(t,t') \Pi_> (t,t')+\Theta_{c}(t',t) \Pi_<(t,t') 
\ea

Now we can do the integrals sequentially using (\ref{eq:matrixmult}).
First we do the integral over $t_2$. Then we write $D_0$ schematically
as
\ba
D_0(t,t') &=& d_0(t,t') - \int_c ~dt_1 U(t,t_1) D_0(t_1,t') \nonumber \\
 &\equiv & d_0(t,t') - \tilde D(t,t')
\ea
where $U(t,t_1)$ is causal and is determined by the matrix
multiplication of $d_0$ and $\Pi$. Then
\ba
\tilde{D} _{ \stackrel{>}{<} }(t,t') & = & - \int_{0}^{t'} dt_3 U_{
\stackrel{>}{<} }(t,t_3) \left[ D_{0 >}(t_3,t') - D_{ 0 <}(t_3,t')
\right]                                                     \nonumber \\ 
&& + \int_{0}^{t} {\rm d}t_3 \left[ U_{>}(t,t_3) - U_{<}(t,t_3)
\right] D _{ 0 \stackrel{>}{<} }(t_3,t')~,              \label{eq:dtild}
\ea
with boundary conditions at $t=t'=0$:
\bq
D_0(0,0) = d_0 (0,0); \hspace{.2in} \tilde{D} (0,0) =0~.
\eq
In order to determine the causal matrix (in time)  $\tilde{D}(t,t')$
one recognizes that for causal Green's functions
\bq
A_> (t,t') = A_<(t',t), \hspace{.2in} A_>(t,t') = - A_>(t',t)^{\ast}~.
\label{causal}
\eq

When doing numerical work, the time integrals are replaced by discrete
sums, with $t=m \epsilon$ and $t'= n\epsilon$ where $\epsilon$ is the
time step. The explicitly causal update is then, for $m \geq n$,
\ba
\tilde{D}_{0>}(m,n)& =& \sum_{k=0}^{m-1} [U_>(m,k) - U_<(m,k)] D_{0>}
(k,n)                                                     \nonumber \\
&-& \sum_{k=0}^{n-1} [U_>(m,k)[ D_{0>} (k,n)-D_{0<} (k,n)]~.
\ea 
Thus starting with the known value of $D_0(0,0)$ we can construct the
entire  causal propagator $D_0(t,t')$  using this and the relations
(\ref{causal}). 

\subsection{Initial conditions}

Particle production in the early universe and particle production by
a classical electric field are two external field problems which admit
a particular type of intial condition. One starts these initial value
problems off with no particles in the appropriate matter field and
with the external field in a ``classical'' state (one where the
expectation value dominates quantum fluctuations). This is also the
situation here with $A$ corresponding to the external field and the
$x_i$ to the modes of the quantum matter field.
 
In line with the above desires for the appropriate initial condition,
we want to enforce $\VEV{x_{0 i}(t)}= \VEV{A_q(t)} = 0$ for all $i$
and values of $t$. This is accomplished by choosing  the initial state
to be an adiabatic vacuum  $ a\ket{0}_{ad} = b\ket{0}_{ad} = 0$. 

Next, we note that solutions of (\ref{eq:fgeqn}),  
\ba
\left\{ { {\rm d}^2  \over  {\rm d} t^2 } +  ( m^2 + \tilde{e}^2
\tilde{A}_0^2(t) ) \right\}  f(t)  & = & 0~,        \nonumber \\
\left\{ { {\rm d}^2  \over {\rm d}t^2 } + \tilde{e}^2  | f(t) |^2
\right\} \, g(t) & = & 0~,                          \label{eq:fgeqnx}
\ea
which automatically obey the Wronskian condition can be written in the
form: 
\ba
f(t) & = & {1\over \sqrt{ 2 \Omega_{f}(t) }} e^{ -i \int_{0}^{t}
\Omega_{f}(t') dt' }~,                         \nonumber \\
g(t) & = & {1\over \sqrt{ 2 \Omega_{f}(t) }} e^{ -i \int_{0}^{t}
\Omega_{g}(t') dt' }~, 
\ea
where $\Omega_{f}(t)$ and $\Omega_{g}(t)$ are solutions of the
nonlinear equations, 
\ba
\half \left( {\ddot{\Omega}_{f} \over \Omega_{f}} \right) + \left(
{\dot{\Omega}_{f} \over \Omega_{f}} \right)^2 + \Omega_{f}^2 & = &
\omega_{f}^2~,                                 \nonumber \\
\half \left( {\ddot{\Omega}_{g} \over \Omega_{g}} \right) +
\left({\dot{\Omega}_{g} \over \Omega_{g}} \right)^2 + \Omega_{g}^2 & =
& \omega_{g}^2~, 
\ea
with
\ba
\omega_{f}^2(t) & = & m^2 + \tilde{e}^2 \tilde{A}_0^2(t)~, \nonumber \\
\omega_{g}^2(t) & = & \tilde{e}^2| f(t) |^2 = { \tilde{e}^2  \over  2
\Omega_{f}(t) }~. 
\ea

We want to match our solutions to asymptotic (adiabatic) Heisenberg
operators.  This will be accomplished by making the choices:
\bq
\Omega_{f}(0)       =  \omega_{f}(0)~;~~
\dot{\Omega}_{f}(0) = \dot{\omega}_{f}(0)~;~~
\Omega_{g}(0)       = \omega_{g}(0)~;~~        
\dot{\Omega}_{g}(0)  =\dot{\omega}_{g}(0)~.
\eq
Finally, the following initial conditions for $f(t)$ and $g(t)$ are
obtained 
\ba
f(0) & = & {1  \over \sqrt{ 2 \omega_{f}(0) } }~; \hspace{.4in} g(0)
=  {1  \over \sqrt{ 2 \omega_{f}(0) } }                   \nonumber \\
\dot{f}(0) & = & - { \dot{\omega}_{f}(0) / 2 \omega_{f}(0) + i
\omega_{f}(0)  \over \sqrt{ 2 \omega_{f}(0) } }~;~~~~ \dot{g}(0)  =  -
{ \dot{\omega}_{g}(0) / 2 \omega_{g}(0) + i \omega_{g}(0)  \over
\sqrt{ 2 \omega_{g}(0) } }~.
\ea
These initial conditions have to be supplemented by the initial values
for $\tilde{A}_0(0)$ and $\dot{\tilde{A}}_0(0)$.

The solution of the $1/N$ correction to $\VEV{x^2(t)}$ will tell us
how the time scale for breakdown of the $1/N$ expansion depends on
$N$. (In certain semiclassical problems, it was found in Ref.
\cite{bib:Berman} that the breakdown time went as $\log N $.) We will
also be able to determine how this time scale is related to the
Lyapunov time scale. The realm of validity of the large $N$ expansion
is determined by comparing to the exact quantum problem which we will
solve as a function of $N$. For the quantum problem we need to supply
the parameters of the initial Gaussian. For the initial data
specified above (adiabatic initial conditions),
\bq
G(t=0) = {1 \over 2 \omega_f(0)}~;~~~~D(t=0)= {1 \over 2 \omega_g(0)}~;
\eq
\bq
\Pi_G(0) = {\dot G(0) \over 4 G(0)} = -{ e^2 A(0) \dot{A}(0)  \over 4
\omega_f^2(0)}~;
\eq
\bq
\Pi_D(0) ={\dot D(0) \over 4D(0)} = -{ \Pi_G(0) \over 2 }~.
\eq
This information determines the real and imaginary part of the width
of the initial wave function for the exact calculation. We have
started numerical simulations of both the $1/N$ expansion and the
exact quantum problem. We hope to be able to present our findings in
the near future.

\appendix
\section{The Closed Time Path Formalism}

In scattering theory one is interested in the probability that an
initial state evolves into a particular final state. The boundary
conditions for the Green's functions for the correlation functions in
that situation are the Feynman ones, and these correlation functions
can be obtained from the conventional path integral formalism which
defines transition elements between states at one time, $t$ (usually
taken to be in the infinite past) to states at another time $t'$ (in
the distant future). If the class of paths is restricted to be the
vacuum configuration at both of its endpoints, then the two states are
the $\vert in \rangle$ and $\langle out \vert$ vacuum states of
scattering theory respectively. The generating functional $Z[J, j]$ of
for the Green's function of scattering theory is the transition matrix
element
\bq 
Z[J,j](t,t') = \langle out,t' \vert in,t \rangle_{J,j}  
\eq 
in the presence of the external sources $J$ and $j$. 

By varying with respect to the external sources we obtain matrix
elements of the Heisenberg field operators between the $\vert in
\rangle$ and $\langle out \vert$ states. For this reason we may refer
to the conventional formulation of the generating functional $Z$ as
the ``in-out'' formalism. The time-ordered Green's functions obtained
in this way necessarily obey Feynman boundary conditions, and these
are the appropriate ones for the calculation of transition
probabilities and cross sections between the $\vert in \rangle$ and
$\langle out \vert$ states. On the other hand the off-diagonal
transition matrix elements of the in-out formalism are completely
inappropriate if what we wish to consider is the time evolution of
physical observables from a given set of initial conditions. The in-out
matrix elements are neither real, nor are their equations of motion
causal at first order in $1/N$, where direct self interactions between
the fields appear for the first time. What we require is a generating
functional for {\it diagonal} matrix elements of field operators with
a corresponding modification of the Feynman boundary conditions on
Green's functions to ensure causal time evolution. This ``in-in''
formalism was developed more than thirty years ago by Schwinger,
Bakshi and Mahanthappa and later by Keldysh, and is called the closed
time path (CTP) method \cite{ref:SchKel}.

The basic idea of the CTP formalism is to take a diagonal matrix
element of the system at a given time $t=0$ and insert a complete set
of states into this matrix element at a different (later) time $t'$.
In this way one can express the original fixed time matrix element as
a product of transition matrix elements from $0$ to $t'$ and the time
reversed (complex conjugate) matrix element from $t'$ to $0$. Since
each term in this product is a transition matrix element of the usual
or time reversed kind, standard path integral representations for each
may be introduced. If the same external source operates in the forward
evolution as the backward one, then the two matrix elements are
precisely complex conjugates of each other, all dependence on the
source drops out and nothing has been gained. However, if the forward
time evolution takes place in the presence of one source $J_+$ but the
reversed time evolution takes place in the presence of a {\it
different} source $J_-$, then the resulting functional is precisely
the generating functional we seek. Indeed (setting $j=0$ and $N=1$
here for simplicity), 
\ba
Z_{in}[J_+,J_-] &\equiv& \int [\cd \J]\langle in \vert \j
\rangle_{J_-} \ \langle \j  \vert in \rangle_{J_+}
                                     \qquad\qquad\nonumber \\ 
&=& \int [\cd \J] \langle in \vert {\ct}^* exp \left[-i
\int_{0}^{t^{\prime}} dt  J_{-}(t) \f (t) \right] \vert \J ,t'\rangle
\times                                          \nonumber\\ 
&&\quad\langle \J ,t' \vert \ct \exp \left[i \int_{0}^{t^{\prime}} dt
J_+ (t) \f(t)\right] \vert in \rangle             \label{CTP} 
\ea
so that, for example,
\bq
{\d W_{in}[J_+,J_-] \over \d J_+(t)}\bigg\vert_{J_+=J_-0} = -{\d
W_{in}[J_+,J_-] \over \d J_-(t)}\bigg\vert_{J_+=J_-=0} = \langle in
\vert \f(t)\vert in \rangle 
\eq
is a true expectation value in the given time-independent Heisenberg
state $\vert in \rangle$. Here $\f(t) =\{A(t),x(t)\}$. Since the time
ordering in Eq. (\ref{CTP}) is forward (denoted by $\ct$) along the
time path from $0$ to $t'$ in the second transition matrix element,
but backward (denoted by $\ct^*$) along the path from $t'$ to $0$ in
the first matrix element, this generating functional receives the name
of the closed time path generating functional. If we deform the
backward and forward directed segments of the path slightly in
opposite directions in the complex $t$ plane, the symbol $\ct_{\cc}$
may be introduced for path ordering along the full closed time
contour, $\cc$, depicted in Fig.3. This deformation of the path
corresponds precisely to opposite $i\e$ prescriptions along the
forward and backward directed segments, which we shall denote by
$\cc_{\pm}$ respectively in the following.  

The doubling of sources, fields and integration contours in the CTP
formalism may seem artificial, but in fact it appears naturally as
soon as one discusses the time evolution not of states in Hilbert
space but of density matrices. Then it is clear that whereas $\vert\
\rangle$ ket states evolve with Hamiltonian $H$, the conjugate
$\langle\ \vert$ bra states evolve with $-H$, and the evolution of the
density matrix requires both. Hence a doubling of all sources and
fields in the functional integral representation of its time evolution
kernel is necessary.  Indeed, it is easy to generalize the functional
in (\ref{CTP}) to the case of an arbitrary initial density matrix
$\r$, by defining 
\ba
 Z \left[ J_{+}, J_{-},\r \right] &\equiv & {\rm Tr} \left\{ \r \left(
\ct^* \exp \left[-i \int_{0}^{t^{\prime}} dt  J_- (t) \f (t)\right]
\right)\left( \ct \exp\left[i \int_{0}^{t^{\prime}} dt J_+ (t)
\f(t)\right] \right) \right\}                 \nonumber \\ 
&=& \int [\cd \vf]  [\cd \vf^{\prime}] [\cd \j] \ \langle \vf \vert \r
\vert \vf^{\prime}\ \rangle\ \langle \vf^{\prime}\vert \ct^* exp
\left[-i \int_{0}^{t^{\prime}} dt J_- (t) \f (t) \right]\vert \j
\rangle                                       \nonumber \\ 
&&\times \langle \j \vert \ct exp \left[i \int_{0}^{t^{\prime}}
dt J_+ (t) \f (t) \right] \vert \vf\rangle~.       \label{Zdens}   
\ea
Variations of this generating function will yield Green's functions in
the state specified by the initial density matrix, {\it i.e.}
expressions of the form, 
\bq 
{\rm Tr}\{ \r  \f(t_{1})  \f(t_{2}) \f(t_{3}) ...\}~. \label{green} 
\eq

Introducing the path integral representation for each transition
matrix element in Eq. (\ref{Zdens}) results in the expression, 
\ba
Z \left[ J_{+}, J_{-},\r \right] &=& \int [\cd \vf]  [\cd\vf^{\prime}]
\ \langle \vf \vert \r \vert \vf^{\prime} \rangle \int [\cd \j]
\int_{\vf}^{\j} [\cd \f_+]\int_{\vf^{\prime}}^{\j } [\cd \f_-]\
\times                                                    \nonumber \\ 
&& \exp \left[i\int_{0}^{\infty} dt  \left(\ L[\f_{+}] -
L[\f_{-}]+J_+\f_+  - J_-  \f_-  \right) \right]\ ,        \nonumber \\ 
\label{CTPgen} 
\ea
where $L$ is the classical Lagrangian functional, and we have taken
the arbitrary future time at which the time path closes $t' \rarr
\infty$. 
 
The double path integral over the fields $\f_+$ and $\f_-$ in
(\ref{CTPgen}) suggests that we introduce a two component
contravariant vector of field variables by 
\ba
\f^a = \left(\begin{array}{c} \f_+ \\ \f_- \end{array}\right)~;
\qquad a=1,2
\ea
with a corresponding two component source vector,
\ba
J^a = \left(\begin{array}{c} J_+ \\ J_- \end{array}\right)~;
\qquad a=1,2\ .
\ea
Because of the minus signs in the exponent of (\ref{CTPgen}), it is
necessary to raise and lower indices in this vector space with a $2
\times 2$ matrix with indefinite signature, namely 
\bq 
c_{ab} = diag \ (+1,-1) = c^{ab} \label{metr}
\eq 
so that, for example  
\bq J^ac_{ab}\F^b = J_+ \f_+ - J_-\f_-~.  
\eq 
These definitions imply that the correlation functions of the theory
will exhibit a matrix structure in the $2\times 2$ space. For
instance, the matrix of connected two point functions in the CTP space
is 
\bq 
G^{ab}(t,t') = {\d^{2} W \over \d J_{a}(t) \d J_{b}(t')} \bigg\vert_{J
=0}~.
\eq  
Explicitly, the components of this $2\times 2$ matrix are   
\ba 
G^{21}(t,t') &\equiv& G_> (t,t') = \ i{\rm Tr}\{\r\ \F(t)
\overline\F(t') \}_{con}~,\nonumber \\  
G^{12}(t,t') &\equiv& G_<(t,t') =  i{\rm Tr}\{\r\ \overline\F(t')
\F(t) \}_{con}~,\nonumber\\
G^{11}(t,t') &=& \ i{\rm Tr}\left\{\r\ \ct[\F (t)
\overline\F(t') ] \right\}_{con} = \q (t,t') G_> (t,t') + \q (t',t)
G_< (t,t')\nonumber~, \\ 
G^{22}(t,t') &=& \ i{\rm Tr}\left\{\r\ \ct^* [\F (t) \overline\F(t')]
\right\}_{con}= \q (t',t) G_> (t,t') + \q (t,t') G_< (t,t')~. \nonumber
\label{matr} 
\ea  
Notice that  
\bq
G^{11}(t,t) = G^{22}(t,t)  
\eq with the usual convention that 
\bq 
\q (t, t) =\half~. 
\eq  

The $2 \times 2$ matrix notation has been discussed extensively in the
literature \cite{ref:SchKel}. However, the development of the CTP
formalism is cleaner, both conceptually and notationally, by returning
to the definition of the generating functional (\ref{Zdens}), and
using the composition rule for transition amplitudes along the closed
time contour in the complex plane. Then we may dispense with the $2
\times 2$ matrix notation altogether, and write simply 
\ba  
\int [\cd \j] \langle \vf^{\prime} \vert \ct^* \exp \left[-i
\int_{0}^{\infty} dt \, J_{-} (t) \f (t) \right] \vert \j \rangle
\langle \j \vert \ct \exp \left[i \int_{0}^{\infty} dt \, J_+ (t)
\f(t)\right] \vert \vf \rangle \qquad                \nonumber \\ 
=\langle \vf^{\prime}  \vert \ct_{\cc} \exp \left[i \int_{\cc} dt \, J
(t) \f(t)\right] \vert \vf \rangle\qquad\qquad\qquad 
\ea 
so that (\ref{Zdens}) may be rewritten more concisely in the CTP
complex path ordered form, 
\ba 
Z_{\cc} \left[ J, \r \right] &= & {\rm Tr} \left\{ \r  \left(
\ct_{\cc} \exp\left[i \int_{\cc} dt J (t) \f (t) \right] \right)
\right\}\qquad\qquad\qquad                             \nonumber \\ 
&=& \int [\cd \vf^1] \int [\cd \vf^2]\ \langle \vf^1 \vert \r \vert
\vf^2\ \rangle \int_{\vf^1}^{\vf^2} [\cd \f] \exp \left[i\int_{\cc} dt
\,\left( L[\f] + J\f\right) \right]~.\\
\label{Zfin}
\ea 

The advantage of this form is that it is identical in structure to the
usual expression for the generating functional in the more familiar
in-out formalism, with the only difference of path ordering according
to the complex time contour $\cc$ replacing the ordinary time ordering
prescription along only $\cc_+$.  Hence, all the functional formalism
of the previous section may be taken over line for line, with only
this modification of complex path ordering in the time
integrations. For example, the propagator function becomes
\ba 
G(t,t')&=&\q_{\cc} (t,t') G_>(t,t') + \q_{\cc} (t',t) G_<(t,t')
\nonumber\\ 
&\equiv & \q_{\cc} (t,t') G^{21}(t,t') + \q_{\cc} (t',t) G^{12}(t,t')
\label{CTPg}  
\ea 
where $\q_{\cc}$ is the CTP complex contour ordered theta function
defined by 
\ba 
\q_{\cc} (t,t') \equiv \left\{ \begin{array}{ll} \q (t,t') &\mbox{for
$t,t'$ both on $\cc_+$}\\ \q (t',t)&\mbox{for $t,t'$ both on
$\cc_-$}\\ $1$ &\mbox{for $t$ on $\cc_-$ , $t'$ on $\cc_+$}\\ $0$
&\mbox{for $t$ on $\cc_+$ , $t'$ on $\cc_-$}\end{array} \right.
\label{CTPt} 
\ea 
With this definition of $G(t,t')$ on the closed time contour, the
Feynman rules are the ordinary ones, and matrix indices are not
required. In integrating over the second half of the contour $\cc_-$
we have only to remember to multiply by an overall negative sign to
take account of the opposite direction of integration, according to
the rule,  
\bq 
\int_{\cc}dt = \int_{0\ \cc_+}^{\infty}dt - \int_{0\
\cc_-}^{\infty} dt~.                             \label{neg} 
\eq  

A second simplification is possible in the form of the generating
functional of (\ref{Zfin}), if we recognize that it is always possible
to express the matrix elements of the density matrix as an exponential
of a polynomial in the fields \cite{bib:chu}:
\bq 
\langle \vf^1 \vert \r \vert \vf^2\ \rangle = \exp
\left[R +  R_a(t_0)\vf^a(t_0) +  R_{ab}(t_0)\vf^a(t_0)\vf^b(t_0) +
\ldots\right]~. 
\eq 
Since any density matrix can be expressed in this form, there is no
loss of generality involved in expressing $\r$ as an exponential. If
we add this exponent to that of the action in (\ref{Zfin}), and
integrate over the two endpoints of the closed time path $\vf^1$ and
$\vf^2$, then the only effect of the non-trivial density matrix $\r$
is to introduce source terms into the path integral for
$Z_{\cc}[J,\r]$ with support {\it only} at the endpoints. This means
that the density matrix can only influence the boundary conditions on
the path integral at $t=0$, where the various coefficient functions
$R_a$, $R_{ab}$, {\it etc.} have the simple interpretations of initial
conditions on the one-point (mean field), two-point (propagator),
functions, {\it etc.} It is clear that the equations of motion for $t
\ne 0$ are not influenced by the presence of these terms at $t_0=0$.
In the special case that the initial density matrix describes a
thermal state, $\r_{\b} = \exp\{-\b H\}$ then the trace over $\r_{\b}$
may be represented as an additional functional integration over fields
along the purely imaginary contour from $t=-i\b$ to $t=0$ traversed
before $\cc_-$ in Fig. 3. In this way the Feynman rules for real time
thermal Green's functions are obtained \cite{sem}. Since we consider
general nonequilibrium initial conditions here we have only the
general expression for the initial $\r$ above and no contour along the
negative imaginary axis in Fig. 3. 

To summarize, we may take over all the results of the usual scattering
theory generating functionals, effective actions, and equations of
motion provided only that we 
\begin{enumerate}
\item substitute the CTP path ordered Green's function(s) (\ref{CTPg})
for the ordinary Feynman propagators in internal lines;
\item integrate over the full closed time contour, $\cc$, according to
(\ref{neg}); and  
\item satisfy the conditions at $t=0$ corresponding to the
initial density matrix $\r$. 
\end{enumerate}

\end{document}